\documentclass[%
 aip,
 jmp,
 amsmath,amssymb,
 reprint,onecolumn
]{revtex4-1}

\usepackage{graphicx}
\usepackage{bm}

\usepackage[utf8]{inputenc}
\usepackage[T1]{fontenc}
\usepackage{mathptmx}
\usepackage{etoolbox}
\usepackage{mathrsfs}

\makeatletter
\def\@email#1#2{%
 \endgroup
 \patchcmd{\titleblock@produce}
  {\frontmatter@RRAPformat}
  {\frontmatter@RRAPformat{\produce@RRAP{*#1\href{mailto:#2}{#2}}}\frontmatter@RRAPformat}
  {}{}
}%
\makeatother

\begin{document}

\title{Gravitational collapse of anisotropic cylindrical shearfree fluids with new exact interior solutions of GR}

\author{Marie-No\"elle C\'el\'erier and Nilton O. Santos}

\address{Laboratoire d'\'etude de l'Univers et des Ph\'enom\`enes Extr\^emes (LUX), Observatoire de Paris, Universit\'e PSL, UMR 8262 CNRS, Sorbonne Universit\'e, 5, Place Jules Janssen, F-92190 Meudon, France}

\date{22 April 2025}

\begin{abstract}
We present a study of shearfree gravitational collapse using cylindrically symmetric spacetimes whose interior is a non-rotating dissipative fluid bounded by a cylindrical hypersurface beyond which is an Einstein-Rosen vacuum exterior. We consider three different pressure configurations: axially, azimuthally, and radially directed, for which we find new exact interior solutions of the field equations. We show that the matching conditions cannot be satisfied by the fluid with radial pressure, while the axial and azimuthal cases with a lapse function depending only on the time coordinate do satisfy these constraints. We derive, for both cases, a sufficient condition for an emission of gravitational radiation from the interior towards the exterior. Therefore we show that, at variance with what happens for spherical symmetry, in the simplified picture of an infinite cylinder of anisotropic shearfree matter, gravitational waves can be emitted during collapsing motion.  

\end{abstract}

\maketitle

\section{Introduction}

In spherical symmetry, gravitational collapse occurs without emission of gravitational waves, owing to Birkhoff's theorem. The next simplest symmetry assumption of cylindrical symmetry has been, therefore, considered in the literature, mostly for dust, see, e. g., \cite{G19}. Now, cylindrical systems may seem physically irrelevant since they impose infinitely long sources. However, under controlled circumstances, they can yield interesting physical descriptions that otherwise might be extremely difficult to tackle mathematically. For instance, this is the case with exact models of spacetime rotation or dragging, models of extragalactic jets, gravitational radiation, translating fluids which might represent light beams produced by stars, and many others. For an overview of the importance of cylindrical systems and cylindrical gravitational waves in General Relativity (GR), see \cite{B20,S24}. 

We establish here the conditions allowing a collapsing cylinder of fluid with anisotropic pressure to emit such waves. The study of shearfree cylindrical gravitational collapse, initiated by Di Prisco et al. \cite{DP09}, is thus completed for cylinders of fluids exhibiting particular kinds of anisotropic pressure.  The shearfree approximation is valid in the limit of very slow collapse. A set of three new exact solutions to the field equations of GR, each corresponding to a principal direction of the fluid pressure, is displayed. However, the matching conditions to an exterior Einstein-Rosen vacuum can only be satisfied for two of them: the spacetimes sourced by a fluid with axial pressure and those sourced by a fluid with azimuthal pressure. 

The translational Killing field $\partial_z$, present in a cylindrical spacetime, as well as the rotational Killing field $\partial_{\phi}$, implies that such spacetimes are not generally asymptotically flat. Therefore, quantities which are usually determined asymptotically, such as mass or energy, are devoid of meaning here. Thus, the definition of radiation for isolated bodies proposed in \cite{B62} cannot be used. 

The conditions to be verified by the parameters defining these solutions to allow the fluid to emit gravitational waves are here obtained by imposing the energy density of the gravitational fluid to decrease with time, while the cylinder is collapsing. We thus identify and expressly quantify the conditions for gravitational waves to occur in a case with only one degree of symmetry less than in the spherical symmetric configuration.

The paper is organized as follows. The metric and general shearfree conditions are displayed in Sec.\ref{shear}. Then, each pressure configuration is examined in turn: the axial case in Sec.\ref{axial}, the azimuthal case in Sec.\ref{azim}, and the radial case in Sec.\ref{rad}. Only the first two are shown to exhibit gravitational radiation under some conditions which are derived there. Section \ref{concl} is devoted to the conclusions.

\section{Metric and shearfree conditions} \label{shear}

The source of the spacetimes is a collapsing cylindrically symmetric anisotropic fluid. It is non-rotating, dissipative and bounded by a cylindrical hypersurface $\Sigma$. In the following, its principal stresses $P_r$, $P_z$ and $P_\phi$ will satisfy three different equations of state which will be analyzed in turn.

In geometric units $c=G=1$, the time dependent diagonal line element reads
\begin{equation}
\textrm{d}s^2=-A^2 (\textrm{d}t^2 - \textrm{d}r^2) + B^2 \textrm{d}z^2 + C^2 \textrm{d}\phi^2, \label{metric}
\end{equation}
$A$, $B$, and $C$, being real functions of the time coordinate $t$ and of the radial coordinate $r$ such as to account for collapse, which can be viewed as an expansion with a reversal of the time coordinate. At some initial time, $t = t_i$, the expansion of the fluid lines is negative and the cylinder starts collapsing. At $t = 0$ the collapse ends in a singularity. Moreover, the cylindrical symmetry forces the coordinates to conform to the following ranges:
\begin{equation}
t_i \geq t > 0, \quad 0 \leq r \leq +\infty, \quad -\infty \leq z \leq +\infty, \quad 0 \leq \phi \leq 2 \pi, \label{ranges}
\end{equation}
where the two limits of the coordinate $\phi$ are topologically identified. The coordinates are denoted $x^0=t$, $x^1=r$, $x^2=z$, and $x^3=\phi$.

We assume that the collapse is slow enough to justify a vanishing shear approximation.

The non-zero components of the shear tensor corresponding to the diagonal metric (\ref{metric}) are \cite{DP09}
\begin{equation}
\sigma_{11} = \frac{A}{3} \left(\frac{2\dot{A}}{A} - \frac{\dot{B}}{B} - \frac{\dot{C}}{C}\right), \label{shear1}
\end{equation}
\begin{equation}
\sigma_{22} = \frac{B^2}{3A} \left(\frac{2\dot{B}}{B} - \frac{\dot{A}}{A} - \frac{\dot{C}}{C}\right), \label{shear2}
\end{equation}
\begin{equation}
\sigma_{33} = \frac{C^2}{3A} \left(\frac{2\dot{C}}{C} - \frac{\dot{A}}{A} - \frac{\dot{B}}{B}\right), \label{shear3}
\end{equation}
where the dots stand for differentiation with respect to $t$. Assuming that the motion of the collapsing fluid is shearfree implies
\begin{equation}
\frac{\dot{A}}{A} = \frac{\dot{B}}{B} = \frac{\dot{C}}{C}, \label{shear4}
\end{equation}
which can be integrated by
\begin{equation}
B = \beta(r) A, \label{shear5}
\end{equation}
\begin{equation}
C = \chi(r) A. \label{shear6}
\end{equation}

\section{Axial pressure} \label{axial}

\subsection{Equations describing the problem} \label{spax}

In this section, $P_r= P_{\phi}=0$, which means that the pressure of the gravitational fluid is axially directed. Its stress-energy tensor, whose general expression has been given by (1) of C\'el\'erier and Santos \cite{CS20}, can therefore be written under the form
\begin{equation}
T_{\alpha \beta} = \rho  V_{\alpha}V_{\beta} + P_z S_\alpha S_\beta, \label{1ax}
\end{equation}
where $\rho$ denotes the energy density of the fluid, $V_\alpha$, its timelike four-velocity, and $S_\alpha$, a spacelike four-vector satisfying
\begin{equation}
V^\alpha V_\alpha = -1, \quad S^\alpha S_\alpha = 1, \quad
 V^\alpha S_\alpha =0. \label{2ax}
\end{equation}

Collapse allows us to choose a frame comoving with the fluid \cite{DP09}. Thus, its four-velocity can be written as
\begin{equation}
V_\alpha = -A \delta_{\alpha}^0, \label{3ax}
\end{equation}
and the spacelike four-vector, as
\begin{equation}
S_{\alpha} = B \delta_{\alpha}^2.  \label{5ax}
\end{equation}

\subsubsection{Field equations} \label{feax}

Using (\ref{3ax}) and (\ref{5ax}) into (\ref{1ax}), the components of the stress-energy tensor corresponding to the five nonvanishing components of the Einstein tensor are obtained, and the five corresponding field equations can be specialized, from their general form as displayed by Di Prisco et al. \cite{DP09}, as

\begin{equation}
G_{00} = \frac{\dot{A}}{A} \left(\frac{\dot{B}}{B} + \frac{\dot{C}}{C}\right) + \frac{\dot{B}}{B}\frac{\dot{C}}{C} - \frac{B''}{B} - \frac{C''}{C} + \frac{A'}{A} \left(\frac{B'}{B} + \frac{C'}{C}\right) - \frac{B'}{B} \frac{C'}{C}= \kappa \rho A^2, \label{G00a}
\end{equation}
\begin{equation}
G_{01} = - \frac{\dot{B}'}{B}  - \frac{\dot{C}'}{C} + \frac{\dot{A}}{A} \left(\frac{B'}{B} + \frac{C'}{C}\right) + \left(\frac{\dot{B}}{B} + \frac{\dot{C}}{C}\right) \frac{A'}{A} = 0, \label{G01a}
\end{equation}
\begin{equation} 
G_{11} = - \frac{\ddot{B}}{B} - \frac{\ddot{C}}{C} + \frac{\dot{A}}{A} \left(\frac{\dot{B}}{B} + \frac{\dot{C}}{C}\right) -  \frac{\dot{B}}{B}\frac{\dot{C}}{C} + \frac{A'}{A} \left(\frac{B'}{B} + \frac{C'}{C}\right) + \frac{B'}{B} \frac{C'}{C} =0, \label{G11a}
\end{equation}
\begin{equation}
G_{22} = \frac{B^2}{A^2}\left( -\frac{\ddot{A}}{A}  -\frac{\ddot{C}}{C} + \frac{\dot{A}^2}{A^2} + \frac{A''}{A} + \frac{C''}{C} - \frac{A'^2}{A^2} \right) = \kappa P_z B^2, \label{G22a}
\end{equation}
\begin{equation}
G_{33} = \frac{C^2}{A^2}\left( -\frac{\ddot{A}}{A}  -\frac{\ddot{B}}{B} + \frac{\dot{A}^2}{A^2} + \frac{A''}{A} + \frac{B''}{B} - \frac{A'^2}{A^2} \right) = 0, \label{G33a}
\end{equation}
where the dots stand still for differentiation with respct to $t$, and the primes, for differentiation with respect to $r$.

\subsubsection{Conservation of the stress-energy tensor} \label{biax}

The conservation of the stress-energy tensor is implemented by the Bianchi identities, whose two non-trivial components have been given as (18) and (19) by Di Prisco et al. \cite{DP09}. Specialized to the present equation of state, they become
\begin{equation}
\dot{\rho} + \rho \left(\frac{\dot{A}}{A} + \frac{\dot{C}}{C}\right) + \left(\rho + P_z\right)\frac{\dot{B}}{B} = 0, \label{Bianchi1ax}
\end{equation}
\begin{equation}
\rho \frac{A'}{A} - P_z \frac{B'}{B} = 0. \label{Bianchi2ax}
\end{equation}

Now, we define the ratio \cite{C21}
\begin{equation}
h = \frac{P_z}{\rho}, \label{hdefax}
\end{equation}
which we insert into (\ref{Bianchi1ax}) and (\ref{Bianchi2ax}) so as to obtain
\begin{equation}
\frac{\dot{\rho}}{\rho} + \frac{\dot{A}}{A} + \frac{\dot{C}}{C} + (1 + h)\frac{\dot{B}}{B} = 0, \label{Bianchi3ax}
\end{equation}
\begin{equation}
\frac{A'}{A} - h \frac{B'}{B} = 0. \label{Bianchi4ax}
\end{equation}

\subsection{Identifying three classes of solutions} \label{three}

We substitute (\ref{shear5}) into (\ref{Bianchi4ax}) and obtain
\begin{equation}
\frac{\beta'}{\beta} = \frac{(1-h)}{h} \frac{A'}{A}, \label{three1}
\end{equation}
from which we derive
\begin{equation}
\frac{\beta''}{\beta} = \frac{(1-h)}{h}\frac{A''}{A} - \frac{h'}{h^2}\frac{A'}{A} +\frac{(1-h)(1-2h)}{h^2}\frac{A'^2}{A^2}. \label{three2}
\end{equation}
Now, using the shearfree conditions in (\ref{three1}), we obtain 
\begin{equation}
\frac{\dot{B}'}{B} = \frac{\dot{A}'}{A} + \frac{(1-h)}{h} \frac{\dot{A} A'}{A^2}. \label{three3}
\end{equation}
Finally, (\ref{three1}) and (\ref{three2}), implemented with the shearfree conditions, give
\begin{equation}
\frac{B''}{B} = \frac{A''}{h A} - \frac{h' A'}{h^2 A} +\frac{(1-h)}{h^2}\frac{A'^2}{A^2}. \label{three4}
\end{equation}

Then, we insert all the above into the Bianchi and the field equations. The first Bianchi identity (\ref{Bianchi3ax}) becomes
\begin{equation}
\frac{\dot{\rho}}{\rho} + (3 + h)\frac{\dot{A}}{A} = 0, \label{Bianchi5}
\end{equation}
while the second Bianchi identity is kept under its form given by (\ref{Bianchi4ax}).
The field equations become
\begin{equation}
\frac{3\dot{A}^2}{A^2} - \frac{(1+h)}{h} \frac{A''}{A} + \frac{h' A'}{h^2 A} - \frac{(1 - h - h^2)}{h^2}\frac{A'^2}{A^2} - \frac{(1+h)}{h} \frac{A' \chi'}{A \chi} - \frac{\chi''}{\chi} = \kappa \rho A^2, \label{G00b}
\end{equation}
\begin{equation}
\frac{\dot{A}'}{A}  - \frac{2\dot{A} A'}{A^2} = 0, \label{G01b}
\end{equation}
\begin{equation} 
- \frac{2 \ddot{A}}{A} + \frac{\dot{A}^2}{A^2} + \frac{(2+h)}{h} \frac{A'^2}{A^2} + \frac{(1+h)}{h} \frac{A' \chi'}{A \chi} =0, \label{G11b}
\end{equation}
\begin{equation}
-\frac{2 \ddot{A}}{A} + \frac{\dot{A}^2}{A^2} + \frac{2 A''}{A} - \frac{A'^2}{A^2} + \frac{2 A' \chi'}{A \chi} + \frac{\chi''}{\chi} = \kappa P_z A^2, \label{G22b}
\end{equation}
\begin{equation}
-\frac{2 \ddot{A}}{A} + \frac{\dot{A}^2}{A^2} + \frac{(1+h)}{h} \frac{A''}{A} - \frac{h' A'}{h^2 A} + \frac{(1 - h - h^2)}{h^2} \frac{A'^2}{A^2} = 0. \label{G33b}
\end{equation}

\hfill

It is easy to see that (\ref{G01b}) possesses three solutions, each defining a given class of spacetimes:

\hfill

- class I corresponds to 
\begin{equation}
A' = 0 \quad \Rightarrow \quad A = a(t), \label{clI}
\end{equation}
with $a$ being a function of the time coordinate $t$ only.

\hfill

- class II corresponds to 
\begin{equation}
\dot{A} = 0 \quad \Rightarrow \quad A = \alpha(r), \label{clII}
\end{equation}
with $\alpha$ being a function of the radial coordinate $r$ only.

\hfill

- class III corresponds to 
\begin{equation}
A = \frac{1}{a(t) + \alpha(r)}. \label{clIII}
\end{equation}

We are therefore going to consider each class of solutions in turn. To facilitate the reading, the functions $a$ and $\alpha$, which are different functions in each class, keep the same names.

\subsection{Class I} \label{I}

\subsubsection{Integration of class I solutions}

We have seen that class I is defined by (\ref{clI}), which we insert into the Bianchi identities so as to obtain
\begin{equation}
\frac{\dot{\rho}}{\rho} + (3 + h)\frac{\dot{a}}{a} = 0, \label{Bianchi5I}
\end{equation}
\begin{equation}
\beta' = 0, \label{Bianchi6I}
\end{equation}
which imply
\begin{equation}
B(t) = c_B a(t), \label{Bianchi7I}
\end{equation}
where $c_B$ is an integration constant which can be absorbed in a rescaling of the $z$ coordinate which yields
\begin{equation}
B(t) = a(t) = A(t). \label{Bianchi8I}
\end{equation}

Now, (\ref{clI}) is substituted similarly into the four remaining field equations. The $G_{00}$ equation thus reads
\begin{equation}
\frac{3\dot{a}^2}{a^2} - \frac{\chi''}{\chi} = \kappa \rho a^2. \label{G00I}
\end{equation}
The $G_{11}$ equation becomes
\begin{equation} 
- \frac{2 \ddot{a}}{a} + \frac{\dot{a}^2}{a^2} = 0, \label{G11I}
\end{equation}
which can be integrated as
\begin{equation}
a(t) = \frac{a_1^2}{4 a_2}t^2 + a_1 t + a_2, \label{I3}
\end{equation}
where $a_1$ and $a_2$ are integration constants. As regards $G_{22}$, we obtain
\begin{equation}
-\frac{2 \ddot{a}}{a} + \frac{\dot{a}^2}{a^2} + \frac{\chi''}{\chi} = \kappa P_z a^2, \label{G22I}
\end{equation}
where we insert (\ref{G11I}), which yields the following expression for the pressure:
\begin{equation}
P_z = \frac{1}{\kappa a^2}\frac{\chi''}{\chi}. \label{I4}
\end{equation}
Finally, $G_{33}$ yields the same equation as $G_{11}$.

Now, (\ref{G00I}) and (\ref{I4}) give, with (\ref{hdefax}) inserted,
\begin{equation}
(1+h) \rho = \frac{3\dot{a}^2}{\kappa a^4}, \label{I5}
\end{equation}
which, with (\ref{Bianchi5I}) inserted, yields
\begin{equation}
\frac{\dot{\rho} a}{\dot{a}} + 2 \rho = - \frac{3\dot{a}^2}{\kappa a^4}. \label{I7}
\end{equation}
We note that the right hand side of this equality is a function of $t$ only, hence, its derivative with respect to $r$ vanishes, which gives
\begin{equation}
\frac{\dot{\rho}'}{\rho'} = - \frac{2\dot{a}}{a}, \label{I8}
\end{equation}
which can be integrated as
\begin{equation}
\rho = \frac{\rho_1(r)}{a^2} + \rho_2(t), \label{I9}
\end{equation}
where $\rho_1$ is a function of $r$ only and $\rho_2$, a function of $t$ only.

Then, we substitute (\ref{I3}) and (\ref{I9}) into (\ref{I7}), which gives the following first order differential equation for $\rho_2$:
\begin{equation}
\dot{\rho_2} + \frac{\frac{a_1^2}{a_2}t + 2 a_1}{\frac{a_1^2}{4 a_2}t^2 + a_1 t + a_2}\rho_2 + \frac{3}{\kappa}\frac{\left(\frac{a_1^2}{2 a_2}t + a_1\right)^3}{\left(\frac{a_1^2}{4 a_2}t^2 + a_1 t + a_2\right)^5} = 0, \label{I11}
\end{equation}
whose solution can be written as
\begin{equation}
\rho_2 = \frac{3\dot{a}^2}{\kappa a^4} + \frac{c_{\rho}}{a^2}, \label{I13}
\end{equation}
with $c_{\rho}$ an integration constant. The pressure follows from (\ref{I5}) as
\begin{equation}
P_z = \frac{3\dot{a}^2}{\kappa a^4} - \rho. \label{I14}
\end{equation}

Now, we insert (\ref{I14}), where we have substituted (\ref{I9}) and (\ref{I13}), into (\ref{I4}) and obtain
\begin{equation}
\rho_1 = - \frac{\chi''}{\kappa \chi} -  c_{\rho}, \label{I15}
\end{equation}
which we substitute into (\ref{I9}), together with (\ref{I13}), and obtain
\begin{equation}
\rho = - \frac{1}{\kappa a^2}\frac{\chi''}{\chi} + \frac{3\dot{a}^2}{\kappa a^4}. \label{I16}
\end{equation}

\subsubsection{Final form of class I solutions} \label{finalI}

The metric reads
\begin{equation}
\textrm{d}s^2=-A^2 \left(\textrm{d}t^2 - \textrm{d}r^2 - \textrm{d}z^2 - \chi ^2\textrm{d}\phi^2\right), \label{metricfinI}
\end{equation}
with
\begin{equation}
A(t) = \frac{a_1^2}{4 a_2}t^2 + a_1 t + a_2 = \frac{a_1^2}{a_2}\left(\frac{t}{2} + \frac{a_2}{a_1}\right)^2. \label{AfinI}
\end{equation}

The energy density and the pressure are
\begin{equation}
\rho(t,r) = - \frac{1}{\kappa a^2}\frac{\chi''}{\chi} + \frac{3\dot{a}^2}{\kappa a^4}, \label{rhofinI}
\end{equation}
\begin{equation}
P_z(t,r) = \frac{1}{\kappa a^2}\frac{\chi''}{\chi}. \label{PfinI}
\end{equation}

\subsubsection{Expansion} \label{expI}

The expansion scalar, as established by Di Prisco et al. \cite{DP09}, reads
\begin{equation}
\Theta = \frac{1}{A}\left(\frac{\dot{A}}{A} + \frac{\dot{B}}{B} +  \frac{\dot{C}}{C}\right) = \frac{3 \dot{A}}{A^2}. \label{expIa}
\end{equation}
By inserting (\ref{AfinI}) into (\ref{expIa}), we obtain
\begin{equation}
\Theta = \frac{3 a_2}{a_1^2 \left(\frac{t}{2} + \frac{a_2}{a_1}\right)^3}. \label{expIb}
\end{equation}

The expansion scalar describes the rate of increase of a volume element of the gravitational fluid
\begin{equation}
\Theta = \frac{1}{V} \frac{\textrm{d}V}{\textrm{d}\tau}, \label{Th1}
\end{equation}
$\tau$ being the proper time of the particle located at the center of the spherical volume element $\textrm{d} V$ \cite{M73}. In a collapse, the volume $V$ decreases as the proper time of the particle increases, therefore the expansion scalar must be negative, which implies
\begin{equation}
\frac{t}{2a_2} + \frac{1}{a_1} < 0. \label{Th2}
\end{equation}
It is easy to see that (\ref{Th2}) can be satisfied by a slowly collapsing fluid, in three cases:

(1) $a_1>0$ and $a_2<0$, hence $t > t_f = -2a_2/a_1>0$, with $t_f<t<t_i$. The collapse begins at $t_i$ and ends at $t_f$.

(2) $a_1<0$ and $a_2>0$, hence $t < t_i = -2a_2/a_1>0$. Equation (\ref{Th2}) is therefore satisfied for $t<t_f$. The collapse still begins at $t_i$ and possibly ends at $t=0$.

(3) $a_1<0$ and $a_2<0$, hence $0<-2a_2/a_1<t$. Since $t$ is always positive, this condition is realized for any allowed value of the time coordinate, which can therefore span its whole definition interval.

\subsubsection{Determination of the $\chi(r)$ function} \label{chiI}

The five independent field equations have allowed us to obtain an integrated class of solutions depending on a function of the radial coordinate, $\chi(r)$. The interpretation of this function comes easily from (\ref{rhofinI}) and (\ref{PfinI}). The ratio $\chi''/\chi$ gives the dependence on $r$ of both the energy density and the axial pressure.

Therefore, a full determination of the metric can follow, for example, from an extra equation of state implying a choice of the ratio $h(t,r)$ defined by (\ref{hdefax}). Indeed, from (\ref{rhofinI}) and (\ref{PfinI}), we have
\begin{equation}
\frac{1+h}{h}(t,r) = H_1(t) H_2(r), \label{I18}
\end{equation}
with
\begin{equation}
H_1(t) = \frac{3 \dot{a}^2}{a^2}, \qquad  H_2(r) = \frac{\chi}{\chi''}. \label{I19}
\end{equation}

The remaining indeterminacy, due to one degree of freedom left in the setting of the problem, is therefore fully encapsulated into $H_2(r)$, and $\chi$ is obtained as a solution of the ordinary second order differential equation
\begin{equation}
H_2(r) {\chi''} - {\chi} = 0. \label{I20}
\end{equation}
Any $H(r)$ function allowing us to integrate this equation yields a class of explicit solutions only determined by a set of constant parameters. 

As a mere example, we can consider the following expression for $H_2$: $H_2(r) = r^2$. This expression satisfies indeed the regularity condition $g_{\phi,\phi} = 0$ on the axis where $r=0$, which condition has to be met by any axially symmetric metric \cite{M93,P93,S09}. It yields, as a solution of (\ref{I20}),
\begin{equation}
\chi  = c_1r^{\frac{1}{2}\left(1+\sqrt{5} \right)} + c_2 r^{\frac{1}{2}\left(1-\sqrt{5} \right)}, \label{I28}
\end{equation}
where $c_1$ and $c_2$ are integration constants. In this case the physical properties of the fluid read
\begin{equation}
h(t,r) = \frac{1}{\frac{3 r^2 \dot{a}^2}{a^2} - 1}, \label{I29}
\end{equation}
\begin{equation}
\rho(t,r) = \frac{1}{\kappa a^2} \left(\frac{3\dot{a}^2}{a^2} - \frac{1}{r^2} \right), \label{I30}
\end{equation}
\begin{equation}
P_z(t,r) = \frac{1}{\kappa a^2 r^2}. \label{I31}
\end{equation}
This implies that the weak energy conditions and the strong energy conditions are both verified in the region of spacetime where
\begin{equation}
\frac{\sqrt{3} r}{|a_2|} > - \left( \frac{t}{2 a_2} + \frac{1}{a_1}\right).
\end{equation}

\subsubsection{Radiation condition} \label{rc}

In a collapsing spacetime, a decrease of the energy density is a sufficient condition for the emission of gravitational waves. We consider the general form of the energy density given by (\ref{rhofinI}) and calculate the partial derivative of $\rho$ with respect to $t$, and obtain, after inserting (\ref{I3}),
\begin{equation}
\frac{\partial \rho}{\partial t} =  \frac{2 a_2^2}{\kappa a_1^4 \left(\frac{t}{2} + \frac{a_2}{a_1}\right)^5}\left[\frac{\chi''}{\chi} - \frac{9}{2 \left(\frac{t}{2} + \frac{a_2}{a_1}\right)^2}\right]. \label{I33}
\end{equation}
Using the notation in (\ref{I20}), we obtain that the sign of $\partial \rho / \partial t$ is produced by the term
\begin{equation}
\left(\frac{t}{2} + \frac{a_2}{a_1}\right)\left[\frac{1}{H_2} - \frac{9}{2 \left(\frac{t}{2} + \frac{a_2}{a_1}\right)^2}\right]. \label{I36}
\end{equation}
In the case where $a_2<0$, we have, from (\ref{Th2}), $\left(\frac{t}{2} + \frac{a_2}{a_1}\right)>0$. Thus, $\rho$ decreases and radiation is emitted, if
\begin{equation}
\frac{1}{H_2} < \frac{9}{2\left(\frac{t}{2} + \frac{a_2}{a_1}\right)^2}. \label{I38a}
\end{equation}
In an interior spacetime defined by the couple of parameters ${a_1,a_2}$, for each cylindrical shell, determined by a given value of $r$, and, thus, by a given value of $H_2(r)$, (\ref{I38a}) must be considered as a constraint on the time coordinate $t$. Therefore, denoting by a lower index $\Sigma$ the values taken on the boundary, radiation is emitted from the interior towards the exterior vacuum if
\begin{equation}
\frac{1}{H_2(r_{\Sigma})} < \frac{9}{2 \left(\frac{t_{\Sigma}}{2} + \frac{a_2}{a_1}\right)^2}. \label{I39}
\end{equation}

In the case where $a_2>0$, $\left(\frac{t}{2} + \frac{a_2}{a_1}\right)$ becomes positive. Thus, $\rho$ decreases and radiation is emitted, if
\begin{equation}
\frac{1}{H_2} > \frac{9}{2\left(\frac{t}{2} + \frac{a_2}{a_1}\right)^2}, \label{I38}
\end{equation}
and at the boundary, we must have
\begin{equation}
\frac{1}{H_2(r_{\Sigma})} > \frac{9}{2 \left(\frac{t_{\Sigma}}{2} + \frac{a_2}{a_1}\right)^2}. \label{I39}
\end{equation}

\subsubsection{Junction condition} \label{junctionax}

We consider the junction conditions to an exterior Einstein-Rosen spacetime \cite{E37} described by the metric
\begin{equation}
\textrm{d}s^2=-\textrm{e}^{2(\gamma - \psi)} \left(\textrm{d}T^2 - \textrm{d}R^2\right) + \textrm{e}^{2\psi}\textrm{d}z^2 +  \textrm{e}^{-2\psi}R^2\textrm{d}\phi^2. \label{metricER}
\end{equation}

Applying Darmois' junction conditions \cite{D27}, Di Prisco et al. \cite{DP09} have obtained, from the continuity of the first fundamental forms through the boundary $\Sigma$, an equality which can be written as
\begin{equation}
\textrm{e}^{2(\gamma - \psi)}\left(\textrm{d}T^2 - \textrm{d}R^2\right) \stackrel{\Sigma}{=} A^2 (\textrm{d}t^2 - \textrm{d}r^2), \label{jcI2}
\end{equation}
and their Eqs. (24) and (25). From the continuity of the second fundamental forms, they have found
\begin{equation}
P_r \stackrel{\Sigma}{=} 0, \label{jcI3}
\end{equation}
and their Eqs. (27) and (28). Then, they have rearranged these equations together with the field equations of the Einstein-Rosen spacetime such as to obtain
\begin{equation}
\textrm{e}^{\psi} \stackrel{\Sigma}{=} B, \label{jcI4}
\end{equation}
\begin{equation}
\psi_{,T} \stackrel{\Sigma}{=} \frac{B_{,t}(BC)_{,r} - B_{,r}(BC)_{,t}}{B\left[(BC)^2_{,r} - (BC)^2_{,t}\right]}, \label{jcI5}
\end{equation}
\begin{equation}
\psi_{,R} \stackrel{\Sigma}{=} \frac{B_{,r}(BC)_{,r} - B_{,t}(BC)_{,t}}{B\left[(BC)^2_{,r} - (BC)^2_{,t}\right]}, \label{jcI6}
\end{equation}
\begin{equation}
\textrm{e}^{\gamma} \stackrel{\Sigma}{=} \frac{AB}{\left[(BC)^2_{,r} - (BC)^2_{,t}\right]^{\frac{1}{2}}}, \label{jcI7}
\end{equation}
\begin{equation}
\gamma_{,T} \stackrel{\Sigma}{=} \frac{2C\left[B_{,t}(BC)_{,r} - B_{,r}(BC)_{,t}\right]\left[B_{,r}(BC)_{,r} - B_{,t}(BC)_{,t}\right]}{B\left[(BC)^2_{,r} - (BC)^2_{,t}\right]^2}, \label{jcI8}
\end{equation}
\begin{equation}
\gamma_{,R} \stackrel{\Sigma}{=} \frac{C\left\{\left[B_{,t}(BC)_{,r} - B_{,r}(BC)_{,t}\right]^2 + \left[B_{,r}(BC)_{,r} - B_{,t}(BC)_{,t}\right]^2\right\}}{B\left[(BC)^2_{,r} - (BC)^2_{,t}\right]^2}, \label{jcI9}
\end{equation}
where $\stackrel{\Sigma}{=}$ denotes that the values are taken at the boundary.

Now, the main junction condition setting constraints on the parameters of the interior spacetimes, independently of the exterior parameters is (\ref{jcI3}). It is already well-known as a junction condition between a rotating cylinder of stationary fluid and a Lewis vacuum \cite{C23b,C23c,C23d,D06}.

Since the equation of state for a fluid with axially directed pressure implies $P_r = 0$ everywhere, this applies also at the boundary and the matching can be completed by imposing the different relations between the parameters of both solutions, the interior class I, and the Einstein-Rosen vacuum exterior. The junction conditions (\ref{jcI4})-(\ref{jcI9}) become, with the interior metric functions describing class I,
\begin{equation}
\textrm{e}^{\psi} \stackrel{\Sigma}{=} A = \frac{a_1^2}{a_2}\left(\frac{t}{2} + \frac{a_2}{a_1}\right)^2, \label{jcI10}
\end{equation}
\begin{equation}
\psi_{,T} \stackrel{\Sigma}{=} \frac{\dot{A} \chi'}{A \left(A^2 \chi'^2 - 4 \dot{A}^2 \chi^2 \right)}, \label{jcI11}
\end{equation}
\begin{equation}
\psi_{,R} \stackrel{\Sigma}{=} \frac{2 \dot{A}^2 \chi}{A^2\left(4\dot{A}^2 \chi^2 - A^2 \chi'^2\right)}, \label{jcI12}
\end{equation}
\begin{equation}
\textrm{e}^{\gamma} \stackrel{\Sigma}{=} \frac{A}{\sqrt{A^2 \chi'^2 - 4 \dot{A}^2 \chi^2}}, \label{jcI13}
\end{equation}
\begin{equation}
\gamma_{,T} \stackrel{\Sigma}{=} - \frac{4 \dot{A}^3 \chi^2 \chi'}{A^2\left(A^2 \chi'^2 - 4 \dot{A}^2 \chi^2 \right)^2}, \label{jcI14}
\end{equation}
\begin{equation}
\gamma_{,R} \stackrel{\Sigma}{=} \frac{\dot{A}^2 \chi \left( A^2 \chi'^2 + 4 \dot{A}^2 \chi^2 \right)}{A^2 \left(4\dot{A}^2 \chi^2 - A^2 \chi'^2\right)^2}. \label{jcI15}
\end{equation}

Now, we study each junction condition in turn.

Whatever the $H_2(r)$ expression, the left-hand side of (\ref{jcI10}) never vanishes, since $(t/2 + a_2/a_1)$ is always positive as we have seen in Sec.\ref{expI}. Moreover, for finite values of the time coordinate, $\textrm{e}^{\psi}$ takes a finite value on the $\Sigma$ boundary.

The other junction conditions depending on $\chi$ and its first derivative, it is easy to see that for both examples considered in Sec.\ref{chiI}, these quantities take finite values for $r= r_{\Sigma}$ on the boundary. Now, by inserting $A$ given by (\ref{AfinI}) into (\ref{jcI11})-(\ref{jcI15}), we obtain, respectively
\begin{equation}
\psi_{,T} \stackrel{\Sigma}{=} \frac{a_2^2 \chi'}{a_1^4 \left(\frac{t}{2} + \frac{a_2}{a_1}\right)^3 \left[\left(\frac{t}{2} + \frac{a_2}{a_1}\right)^2\chi'^2 - 4 \chi^2 \right]}, \label{jcI16}
\end{equation}
\begin{equation}
\psi_{,R} \stackrel{\Sigma}{=} \frac{2 a_2^2 \chi}{a_1^4 \left(\frac{t}{2} + \frac{a_2}{a_1}\right)^4 \left[4 \chi^2 - \left(\frac{t}{2} + \frac{a_2}{a_1}\right)^2 \chi'^2\right]}, \label{jcI17}
\end{equation}
\begin{equation}
\textrm{e}^{\gamma} \stackrel{\Sigma}{=} \frac{\frac{t}{2} + \frac{a_2}{a_1}}{ \sqrt{\left(\frac{t}{2} + \frac{a_2}{a_1}\right)^2\chi'^2 - 4 \chi^2}}. \label{jcI18}
\end{equation}
\begin{equation}
\gamma_{,T} \stackrel{\Sigma}{=} \frac{4 a_2^3 \chi^2 \chi'}{a_1^6 \left(\frac{t}{2} + \frac{a_2}{a_1}\right)^5 \left[4 \chi^2 - \left(\frac{t}{2} + \frac{a_2}{a_1}\right)^2 \chi'^2\right]^2}, \label{jcI19}
\end{equation}
\begin{equation}
\gamma_{,R} \stackrel{\Sigma}{=} \frac{a_2^2 \left[ 4 \chi^2 + \left(\frac{t}{2} + \frac{a_2}{a_1}\right)^2 \chi'^2 \right]}{a_1^4 \left(\frac{t}{2} + \frac{a_2}{a_1}\right)^4 \left[4 \chi^2 - \left(\frac{t}{2} + \frac{a_2}{a_1}\right)^2 \chi'^2\right]^2}. \label{jcI20}
\end{equation}
Indeed, all the above expressions take finite values on $\Sigma$. We have shown, therefore, that the whole set of junction conditions can be satisfied on any boundary defined by a non-vanishing and non-diverging radial coordinate, for class I solutions describing interior spacetimes.

To exemplify this statement, we apply it to a case where the cylindrical wave starts traveling outward as a short-duration disturbance. Such a solution of the wave-like field equation of Einstein-Rosen spacetimes can be written as \cite{R54,Ca82}
\begin{equation}
\psi = \frac{1}{2 \pi} \int_{- \infty}^{T-R} \frac{f(t') \textrm{d}t'}{\left[(T-t')^2 - R^2\right]^{1/2}}. \label{jcI21}
\end{equation}
For the pulse function, we can choose \cite{R54,Ca82}
\begin{equation}
f(t') = f_0 \delta (t'), \label{jcI22}
\end{equation}
where $f_0$ is a constant and $\delta(t')$ is the Dirac delta distribution. By inserting (\ref{jcI22}) into (\ref{jcI21}), we obtain
\begin{equation}
\psi = 0, \qquad T - R<0, \label{jcI23}
\end{equation}
\begin{equation}
\psi = \frac{f_0}{2 \pi (T^2 - R^2)^{1/2}}, \qquad T - R>0. \label{jcI24}
\end{equation}
By implementing (\ref{jcI23}) and (\ref{jcI24}) into the remaining field equations and solving them, as done in \cite{R54,Ca82}, we obtain
\begin{equation}
\gamma = 0, \qquad T - R<0, \label{jcI25}
\end{equation}
\begin{equation}
\gamma = \frac{f_0^2}{8 \pi^2} \frac{R^2} {(T^2 - R^2)^2}, \qquad T - R>0. \label{jcI26}
\end{equation}

First, we consider the case where $T-R>0$. Thus, (\ref{jcI24}) inserted into (\ref{jcI10})-(\ref{jcI12}) yields, respectively,
\begin{equation}
\textrm{exp} \left[\frac{f_0}{2 \pi (T^2 - R^2)^{1/2}}\right] \stackrel{\Sigma}{=} \frac{a_1^2}{a_2} \left(\frac{t}{2} + \frac{a_2}{a_1}\right)^2, \label{jcI27}
\end{equation}
\begin{equation}
-\frac{f_0}{2 \pi}\frac{T}{(T^2 - R^2)^{3/2}} \stackrel{\Sigma}{=} \frac{a_2^2 \chi'}{2 a_1^4 \left(\frac{t}{2} + \frac{a_2}{a_1}\right)^4 \chi \left[ \left(\frac{t}{2}+\frac{a_2}{a_1}\right)\chi' - 2\chi\right]}, \label{jcI28}
\end{equation}
\begin{equation}
\frac{f_0}{2 \pi}\frac{R}{(T^2 - R^2)^{3/2}} \stackrel{\Sigma}{=} \frac{a_2^2}{a_1^4\left(\frac{t}{2} + \frac{a_2}{a_1}\right)^5 \left[ 2 \chi - \left(\frac{t}{2}+\frac{a_2}{a_1}\right)\chi'\right]}. \label{jcI29}
\end{equation}
Then, by inserting (\ref{jcI26}) into (\ref{jcI13})-(\ref{jcI15}), we obtain, respectively,
\begin{equation}
\textrm{exp} \left[\frac{f_0^2}{8 \pi^2}\frac{R^2}{(T^2 - R^2)^2}\right] \stackrel{\Sigma}{=} 
\left\{\chi \sqrt{2\left[\frac{\chi'}{\chi} - \frac{2}{\left(\frac{t}{2} + \frac{a_2}{a_1}\right)} \right]}\right\}^{-1}, \label{jcI30}
\end{equation}
\begin{equation}
-\frac{f_0^2}{2 \pi^2}\frac{T R^2}{(T^2 - R^2)^3} \stackrel{\Sigma}{=} \frac{a_2^2 \chi'}{a_1^4 \left(\frac{t}{2} + \frac{a_2}{a_1}\right)^5 \left[2\chi - \left(\frac{t}{2}+\frac{a_2}{a_1}\right)\chi' \right]^2}, \label{jcI31}
\end{equation}
\begin{equation}
\frac{f_0^2}{4 \pi^2}\frac{R(T^2 + R^2)}{(T^2 - R^2)^3} \stackrel{\Sigma}{=} \frac{a_1^2 \chi \left[ \left(\frac{t}{2}+\frac{a_2}{a_1}\right)^2 + 4\chi\right]}{a_2 \left[2\chi - \left(\frac{t}{2}+\frac{a_2}{a_1}\right)\chi' \right]^2}. \label{jcI32}
\end{equation}
After some arrangements, we obtain
\begin{equation}
\frac{f_0}{2 \pi}\frac{1}{(T^2 - R^2)^{3/2}} \stackrel{\Sigma}{=} \frac{a_2^7\left[\left(\frac{t}{2} + \frac{a_2}{a_1}\right)^2 \chi'^2 + 4 \chi^2\right]}{a_1^{14} \left(\frac{t}{2} + \frac{a_2}{a_1}\right)^{15} 4 \chi^3 \left[2\chi - \left(\frac{t}{2}+\frac{a_2}{a_1}\right)\chi' \right]  \left[ \left(\frac{t}{2}+\frac{a_2}{a_1}\right)^2 + 4\chi\right]}, \label{jcI33}
\end{equation}
\begin{equation}
T \stackrel{\Sigma}{=} - \frac{a_1^4}{a_2^2}\left(\frac{t}{2} + \frac{a_2}{a_1}\right)^5 \frac{\chi'}{2}, \label{jcI34}
\end{equation}
\begin{equation}
\chi'^2 \stackrel{\Sigma}{=} - 4 \chi^2 \left\{\frac{a_1^6}{a_2^3}\left[ \left(\frac{t}{2}+\frac{a_2}{a_1}\right)^2 + 4\chi\right]\left(\frac{t}{2} + \frac{a_2}{a_1}\right)^4 +\frac{1}{\left(\frac{t}{2} + \frac{a_2}{a_1}\right)^2} \right\}, \label{jcI35}
\end{equation}
\begin{equation}
R \stackrel{\Sigma}{=} - \frac{a_1^4}{a_2^2}\left(\frac{t}{2} + \frac{a_2}{a_1}\right)^4 \chi. \label{jcI36}
\end{equation}

Therefore, we have six independent equations, i. e., (\ref{jcI27}), (\ref{jcI30}), and (\ref{jcI33})-(\ref{jcI36}), for eight parameters, i. e., $a_1$, $a_2$, $f_0$, $\tau_{\Sigma}$, $R_{\Sigma}$, $t_{\Sigma}$, $\chi_{\Sigma}$, and $\chi'_{\Sigma}$. We can thus, by choosing the values of two parameters, deduce those of the six remaining ones. As an example, fixing $a_1$ and $a_2$ allows us to obtain a given value of the strength of the pulse, $f_0$, the location of the boundary in the exterior spacetime, given by $\tau_{\Sigma}$ and $R_{\Sigma}$, its location in the interior spacetime, given by $t_{\Sigma}$, $\chi_{\Sigma}$, and $\chi'_{\Sigma}$, the last two allowing us, in principle, to derive $r_{\Sigma}$.

Finally, in the case where $T-R<0$, an analogous and straightforward reasoning leads to the conclusion that a proper matching of a class I interior to an Einstein-Rosen exterior vacuum is impossible in this spacetime region. This implies that the boundary between both solutions must be located in the region where $T-R>0$.

\subsection{Class II} \label{II}

Class II is defined by 
\begin{equation}
\dot{A} = 0 \quad \Rightarrow \quad A = \alpha(r), \label{II1}
\end{equation}
which we insert into (\ref{shear4}) and obtain
\begin{equation}
\dot{A} = \dot{B} = \dot{C} = 0, \label{II2}
\end{equation}
which we substitute into the Bianchi identity (\ref{Bianchi3ax}) which becomes
\begin{equation}
\dot{\rho} = 0. \label{II3}
\end{equation}

Therefore, the solution is stationary and cannot represent a collapse.

\subsection{Class III} \label{III}

Class III is defined by
\begin{equation}
A = \frac{1}{a(t) + \alpha(r)}. \label{III1}
\end{equation}
The second Bianchi identity (\ref{Bianchi2ax}), becomes
\begin{equation}
\frac{\beta'}{\beta} = - \frac{\alpha'}{h(a + \alpha)}, \label{III7}
\end{equation}
from which we derive
\begin{equation}
\frac{\beta''}{\beta} = - \frac{\alpha''}{h(a + \alpha)} + \frac{h' \alpha'}{h^2(a + \alpha)} + \frac{(1+h)}{h^2}  \frac{\alpha'^2}{(a + \alpha)^2}. \label{III8}
\end{equation}
Moreover, (\ref{III7}) can be written as
\begin{equation}
h = - \frac{H(r)}{a + \alpha}, \label{III9}
\end{equation}
where we define $H(r)$, function of $r$ only, by
\begin{equation}
H = \frac{\alpha' \beta}{\beta'}. \label{III10}
\end{equation}

Now, we consider the field equations which we arrange so as to obtain
\begin{equation}
A = \frac{H \left(\frac{\chi'}{\chi} - \frac{\alpha''}{\alpha'}\right) + H' - \alpha'}{H^2\left(\frac{\chi'}{\chi} - \frac{\alpha''}{\alpha'}\right) + 2 \alpha' H}, \label{III14}
\end{equation}

The right-hand side of (\ref{III14}) is an expression depending only on $r$. Therefore we can write
\begin{equation}
\dot{A} = 0, \label{III15}
\end{equation}
which is the defining equation of class II. We have thus shown that classes II and III are equivalent, i. e., stationary, and, therefore, they cannot represent a collapse.

\section{Azimuthal pressure} \label{azim}

\subsection{Equations describing the problem} \label{spaz}

The gravitational source is still a cylindrically symmetric anisotropic fluid in collapsing motion. It is still non rotating, dissipative and bounded by a cylindrical hypersurface $\Sigma$. The difference with the fluid studied in Sec.\ref{axial} is that its principal stresses $P_r$, $P_z$ and $P_\phi$ satisfy the equation of state $P_r= P_z=0$, which means that its pressure is azimuthally directed. Its stress-energy tensor, issued from the general expression which has been given by (1) of C\'el\'erier and Santos \cite{CS20}, can therefore be written under the form
\begin{equation}
T_{\alpha \beta} = \rho  V_{\alpha}V_{\beta} + P_{\phi} K_\alpha K_\beta, \label{1az}
\end{equation}
where $\rho$ still denotes the energy density of the fluid, $V_\alpha$, its timelike four-velocity, and $K_\alpha$ is a spacelike four-vector satisfying
\begin{equation}
V^\alpha V_\alpha = -1, \quad K^\alpha K_\alpha = 1, \quad
 V^\alpha K_\alpha =0. \label{2az}
\end{equation}
In geometric units $c=G=1$, the time dependent diagonal line element is still given by (\ref{3ax}), and the coordinates conform to the ranges given by (\ref{ranges}).

Collapse still allows the choice of a frame comoving with the fluid \cite{DP09}. Thus, its four-velocity can be written as
\begin{equation}
V_{\alpha} = -A \delta_{\alpha}^0. \label{3az}
\end{equation}

The spacelike four-vector is
\begin{equation}
K_{\alpha} = C \delta_{\alpha}^3.  \label{5az}
\end{equation}

\subsubsection{Field equations} \label{feaz}

Using (\ref{3az}) and (\ref{5az}) into (\ref{1az}), the components of the stress-energy tensor matching the five nonvanishing components of the Einstein tensor are obtained, and the five corresponding field equations can be specialized, from their general form as displayed by Di Prisco et al. \cite{DP09}, as

\begin{equation}
G_{00} = \frac{\dot{A}}{A} \left(\frac{\dot{B}}{B} + \frac{\dot{C}}{C}\right) + \frac{\dot{B}}{B}\frac{\dot{C}}{C} - \frac{B''}{B} - \frac{C''}{C} + \frac{A'}{A} \left(\frac{B'}{B} + \frac{C'}{C}\right) - \frac{B'}{B} \frac{C'}{C}= \kappa\rho A^2, \label{G00az}
\end{equation}
\begin{equation}
G_{01} = - \frac{\dot{B}'}{B}  - \frac{\dot{C}'}{C} + \frac{\dot{A}}{A} \left(\frac{B'}{B} + \frac{C'}{C}\right) + \left(\frac{\dot{B}}{B} + \frac{\dot{C}}{C}\right) \frac{A'}{A} = 0, \label{G01az}
\end{equation}
\begin{equation} 
G_{11} = - \frac{\ddot{B}}{B} - \frac{\ddot{C}}{C} + \frac{\dot{A}}{A} \left(\frac{\dot{B}}{B} + \frac{\dot{C}}{C}\right) -  \frac{\dot{B}}{B}\frac{\dot{C}}{C} + \frac{A'}{A} \left(\frac{B'}{B} + \frac{C'}{C}\right) + \frac{B'}{B} \frac{C'}{C} =0, \label{G11az}
\end{equation}
\begin{equation}
G_{22} = \frac{B^2}{A^2}\left( -\frac{\ddot{A}}{A}  -\frac{\ddot{C}}{C} + \frac{\dot{A}^2}{A^2} + \frac{A''}{A} + \frac{C''}{C} - \frac{A'^2}{A^2} \right) = 0, \label{G22az1}
\end{equation}
which can be written as
\begin{equation}
-\frac{\ddot{A}}{A}  -\frac{\ddot{C}}{C} + \frac{\dot{A}^2}{A^2} + \frac{A''}{A} + \frac{C''}{C} - \frac{A'^2}{A^2} = 0, \label{G22az2}
\end{equation}
\begin{equation}
G_{33} = \frac{C^2}{A^2}\left( -\frac{\ddot{A}}{A}  -\frac{\ddot{B}}{B} + \frac{\dot{A}^2}{A^2} + \frac{A''}{A} + \frac{B''}{B} - \frac{A'^2}{A^2} \right) = \kappa P_{\phi} C^2, \label{G33az1}
\end{equation}
which can be written as
\begin{equation}
-\frac{\ddot{A}}{A}  -\frac{\ddot{B}}{B} + \frac{\dot{A}^2}{A^2} + \frac{A''}{A} + \frac{B''}{B} - \frac{A'^2}{A^2} = \kappa P_{\phi} A^2. \label{G33az2}
\end{equation}

\subsubsection{Conservation of the stress-energy tensor} \label{biaz}

The conservation of the stress-energy tensor is implemented by the Bianchi identities, whose two non trivial components have been given as (18) and (19) by Di Prisco et al. \cite{DP09}. Specialized to the present equation of state, they become
\begin{equation}
\dot{\rho} + \rho \left(\frac{\dot{A}}{A} + \frac{\dot{B}}{B}\right) + \left(\rho + P_{\phi}\right)\frac{\dot{C}}{C} = 0, \label{Bianchi1az}
\end{equation}
\begin{equation}
\rho \frac{A'}{A} - P_{\phi} \frac{C'}{C} = 0. \label{Bianchi2az}
\end{equation}

\subsection{Solving the field equations}

We still assume shearfree motion.

It is therefore easy to remark that all the equations defining the azimuthal pressure configuration can be strictly derived from those occurring in the axial pressure case by switching $B$ and $C$ and $P_z$ and $P_{\phi}$. Therefore, the solutions can be directly derived from the results obtained in this axial pressure case.

We recover three classes issuing from the same $G_{01}$ field equation which we name here:

\hfill

- class A corresponding to 

\hfill

$A' = 0 \quad \Rightarrow \quad A = a(t)$, 

\hfill

with $a$ being a function of the time coordinate $t$ only.

\hfill

- class B corresponding to

\hfill

$\dot{A} = 0 \quad \Rightarrow \quad A = \alpha(r)$,

\hfill

with $\alpha$ being a function of the radial coordinate $r$ only.

\hfill

- class C corresponding to 

\hfill

$A = \frac{1}{a(t) + \alpha(r)}$.

\hfill

\subsubsection{Class A}

The final form of the solutions, as derived from that of the axial pressure case, reads, for the metric
\begin{equation}
\textrm{d}s^2=-A^2 \left(\textrm{d}t^2 - \textrm{d}r^2 - \beta^2 \textrm{d}z^2 - \textrm{d}\phi^2\right), \label{metricfinA}
\end{equation}
with
\begin{equation}
A(t) = \frac{a_1^2}{a_2}\left(\frac{t}{2} + \frac{a_2}{a_1}\right)^2. \label{AfinA}
\end{equation}

The energy density and the pressure are
\begin{equation}
\rho(t,r) = - \frac{1}{\kappa a^2}\frac{\beta''}{\beta} + \frac{3\dot{a}^2}{\kappa a^4}, \label{A16}
\end{equation}
\begin{equation}
P_{\phi}(t,r) = \frac{1}{\kappa a^2}\frac{\beta''}{\beta}. \label{A17}
\end{equation}

{\it a. Radiation condition} In this case, the energy density given by (\ref{A16}) has the same form as (\ref{rhofinI}) belonging to the axial pressure class I, where $\chi$ has been replaced by $\beta$. Therefore, a sufficient condition for a decreasing energy density, thus for radiation emitted from the interior towards the exterior vacuum, is analogous to (\ref{I39}), with $H_2(r) = \beta''/\beta$.

{\it b. Junction conditions} Since the equation of state for a fluid with an azimuthally directed pressure implies $P_r = 0$ everywhere, the main junction condition established by Di Prisco et al. \cite{DP09}, i. e., $P_r \stackrel{\Sigma}{=} 0$, is satisfied trivially again here.

Then, the matching can be completed by imposing the different relations between the parameters of both solutions, the interior class A, and the Einstein-Rosen vacuum exterior.

These other junction conditions have been considered in Sec.\ref{junctionax} for the case of an axially directed pressure. We have shown that the whole set of conditions can be satisfied on any boundary defined by a non-vanishing and non-diverging radial coordinate and with a positive time coordinate, for class I solutions describing the interior spacetimes.  These results can be applied here by switching $B$ and $C$ and $P_z$ and $P_{\phi}$. We obtain thus similar results and a genuine matching of class A solutions to the Einstein-Rosen vacuum exterior.

\subsection{Class B} \label{B}

Class B is defined by 
\begin{equation}
\dot{A} = 0 \quad \Rightarrow \quad A = \alpha(r), \label{B1}
\end{equation}
which, by symmetry between the $B$ and $C$ metric functions gives
\begin{equation}
\dot{\rho} = 0. \label{B2}
\end{equation}

As regards the axial pressure case, the solution is therefore stationary and cannot represent a collapse.

\subsection{Class C} \label{C}

Class C is defined by
\begin{equation}
A = \frac{1}{a(t) + \alpha(r)}. \label{C1}
\end{equation}
which becomes, as shown in Sec.\ref{axial}, with $\beta(r)$ replacing $\chi(r)$ in the reasoning, 
\begin{equation}
\dot{A} = 0. \label{B1}
\end{equation}
As for the axial pressure case, the solution is therefore stationary and cannot represent a collapse.

\section{Radial pressure} \label{rad}

\subsection{Equations describing the problem} \label{spr}

The gravitational system is still a cylindrically symmetric anisotropic fluid in  collapsing motion. It is non rotating, dissipative and bounded by a cylindrical hypersurface $\Sigma$. Its principal stresses $P_r$, $P_z$ and $P_\phi$ satisfy, now, the equation of state $P_z= P_{\phi}=0$, which means that its pressure is radially-directed. Its stress-energy tensor can therefore be written under the form
\begin{equation}
T_{\alpha \beta} = \rho  V_{\alpha}V_{\beta} + P_r (g_{\alpha \beta} + V_{\alpha}V_{\beta} - S_\alpha S_\beta - K_\alpha K_\beta), \label{1r}
\end{equation}
where $\rho$ still denotes the energy density of the fluid, $V_\alpha$, its timelike four-velocity, and $S_\alpha$ and $K_\alpha$, spacelike four-vectors satisfying
\begin{equation}
V^\alpha V_\alpha = -1, \quad K^\alpha K_\alpha = S^\alpha S_\alpha = 1, \quad V^\alpha K_\alpha = V^\alpha S_\alpha = K^\alpha S_\alpha =0. \label{fourvecr}
\end{equation}
In geometric units $c=G=1$, the time dependent diagonal line element still reads as (\ref{metric}), and the  cylindrical symmetry still forces the coordinates to conform to the ranges given by (\ref{ranges}).

Collapse allows us to choose a frame comoving with the fluid \cite{DP09}. Thus, in the case of radial pressure, its four-velocity and the spacelike four-vectors can be written as
\begin{equation}
V_\alpha = -A \delta_{\alpha}^0,\quad S_\alpha = B \delta_{\alpha}^2, \quad K_\alpha = C \delta_{\alpha}^3. \label{3r}
\end{equation}

\subsubsection{Shearfree assumption} \label{shearr}

We still assume that the motion of the collapsing fluid is shearfree. Therefore, the identities obtained from this assumption are the same as in Sec.\ref{shear}.

\subsubsection{Conservation of the stress-energy tensor} \label{bir}

The conservation of the stress-energy tensor is implemented by the Bianchi identities. Specialized to the present equation of state, they read
\begin{equation}
\dot{\rho} + (\rho + P_r) \frac{\dot{A}}{A} + \rho\left(\frac{\dot{B}}{B} +  \frac{\dot{C}}{C}\right) = 0, \label{Bianchi1r}
\end{equation}
\begin{equation}
P'_r + (\rho + P_r)\frac{A'}{A} + P_r \left(\frac{B'}{B} + \frac{C'}{C}\right) = 0. \label{Bianchi2r}
\end{equation}

Now, we define the ratio $h$ by
\begin{equation}
h = \frac{P_r}{\rho}, \label{hdefr}
\end{equation}
which we insert into (\ref{Bianchi1r}) and (\ref{Bianchi2r}) so as to obtain, with (\ref{shear4})-(\ref{shear6}) implemented,
\begin{equation}
\frac{\dot{\rho}}{\rho} + (3 + h)\frac{\dot{A}}{A} = 0, \label{Bianchi3r}
\end{equation}
\begin{equation}
\frac{P'_r}{P_r} + \frac{(1+3h)}{h}\frac{A'}{A} + \frac{\beta'}{\beta} + \frac{\chi'}{\chi}= 0. \label{Bianchi4r}
\end{equation}
Notice that, here, we have kept the same notation for the ratio $h$ while it has a different meaning from that of the $h$ ratio used in the axial pressure case where it plays however an analogous role.

\subsubsection{Field equations} \label{fer}

Using (\ref{3r}) into (\ref{1r}), the components of the stress-energy tensor corresponding to the five nonvanishing components of the Einstein tensor are obtained, and the five corresponding field equations can be specialized to

\begin{equation}
G_{00} = \frac{\dot{A}}{A} \left(\frac{\dot{B}}{B} + \frac{\dot{C}}{C}\right) + \frac{\dot{B}}{B}\frac{\dot{C}}{C} - \frac{B''}{B} - \frac{C''}{C} + \frac{A'}{A} \left(\frac{B'}{B} + \frac{C'}{C}\right) - \frac{B'}{B} \frac{C'}{C}= \kappa\rho A^2, \label{G00r}
\end{equation}
\begin{equation}
G_{01} = - \frac{\dot{B}'}{B}  - \frac{\dot{C}'}{C} + \frac{\dot{A}}{A} \left(\frac{B'}{B} + \frac{C'}{C}\right) + \left(\frac{\dot{B}}{B} + \frac{\dot{C}}{C}\right) \frac{A'}{A} = 0, \label{G01r}
\end{equation}
\begin{equation} 
G_{11} = - \frac{\ddot{B}}{B} - \frac{\ddot{C}}{C} + \frac{\dot{A}}{A} \left(\frac{\dot{B}}{B} + \frac{\dot{C}}{C}\right) -  \frac{\dot{B}}{B}\frac{\dot{C}}{C} + \frac{A'}{A} \left(\frac{B'}{B} + \frac{C'}{C}\right) + \frac{B'}{B} \frac{C'}{C} =\kappa P_r A^2, \label{G11r}
\end{equation}
\begin{equation}
G_{22} = \frac{B^2}{A^2}\left( -\frac{\ddot{A}}{A}  -\frac{\ddot{C}}{C} + \frac{\dot{A}^2}{A^2} + \frac{A''}{A} + \frac{C''}{C} - \frac{A'^2}{A^2} \right) = 0, \label{G22r}
\end{equation}
\begin{equation}
G_{33} = \frac{C^2}{A^2}\left( -\frac{\ddot{A}}{A}  -\frac{\ddot{B}}{B} + \frac{\dot{A}^2}{A^2} + \frac{A''}{A} + \frac{B''}{B} - \frac{A'^2}{A^2} \right) = 0. \label{G33r}
\end{equation}

\subsection{Identifying three classes of solutions} \label{threer}

We divide (\ref{G22r}) by $B^2/A^2$, and (\ref{G33r}) by $C^2/A^2$ and subtract the results which gives
\begin{equation}
\frac{B''}{B} = \frac{C''}{C}, \label{three1r}
\end{equation}
where we insert (\ref{shear5}) and (\ref{shear6}) so as to obtain
\begin{equation}
\frac{A'}{A} = \frac{\frac{\beta''}{\beta} -  \frac{\chi''}{\chi}}{2\left(\frac{\chi'}{\chi} - \frac{\beta'}{\beta}\right)}. \label{three2r}
\end{equation}
Since the left hand side of (\ref{three2r}) is a function of $r$ only, we can write
\begin{equation}
\left(\frac{A'}{A}\right)^{\dot{}} = 0, \label{three3r}
\end{equation}
which can be written as
\begin{equation}
A \dot{A}' - \dot{A} A' = 0. \label{three4r}
\end{equation}

\hfill

It is easy to see that (\ref{three4r}) possesses three solutions, each defining a given class of spacetimes:

\hfill

- class 1 corresponds to 

\hfill

$A' = 0 \quad \Rightarrow \quad A = a(t)$,

\hfill

with $a$ being a function of the time coordinate $t$ only.

\hfill

- class 2 corresponds to 

\hfill

$\dot{A} = 0 \quad \Rightarrow \quad A = \alpha(r)$,

\hfill

with $\alpha$ being a function of the radial coordinate $r$ only.

\hfill

- class 3 corresponds to 

\hfill

$A = \lambda \textrm{e}^{b_1 t + b_2 r}$,

\hfill

where $\lambda$, $b_1$, and $b_2$ are integration constant. We are therefore going to consider each class of solutions in turn. For an easy reading, the functions $a$, $\alpha$, $\beta$, and $\chi$, which are different functions in each class, keep the same notations.

\subsection{Class 1} \label{class1}

We have seen that class 1 is defined by 
\begin{equation}
A' = 0 \quad \Rightarrow \quad A = a(t), \label{1.1}
\end{equation}
which we insert into the Bianchi identities so as to obtain
\begin{equation}
\frac{\dot{\rho}}{\rho} + (3 + h)\frac{\dot{a}}{a} = 0, \label{Bianchi5.1}
\end{equation}
and
\begin{equation}
\frac{P'_r}{P_r} +  \frac{\beta'}{\beta} + \frac{\chi'}{\chi} = 0, \label{Bianchi6.1}
\end{equation}
which implies, since $\frac{\beta'}{\beta} + \frac{\chi'}{\chi}$ is a function of $r$ only
\begin{equation}
P_r(t,r) = P_1(r) P_2(t), \label{Bianchi7.1}
\end{equation}
where $P_1$ is a function of $r$ only and $P_2$, a function of $t$ only. By substituting (\ref{Bianchi7.1}) into (\ref{Bianchi6.1}), we obtain the first order ordinary differential equation
\begin{equation}
\frac{P'_1}{P_1} +  \frac{\beta'}{\beta} + \frac{\chi'}{\chi} = 0, \label{1.3}
\end{equation}
which can be integrated by
\begin{equation}
P_1 =  \frac{c_p}{\beta \chi}, \label{1.4}
\end{equation}
with $c_p$ an integration constant. Then, (\ref{1.4}) substituted into (\ref{Bianchi7.1}) yields
\begin{equation}
P_r(t,r) =  \frac{P_2(t)}{\beta(r) \chi(r)}, \label{1.5}
\end{equation}
where $c_p$ has been absorbed into $P_2(t)$.

Now, the above relations are substituted into the field equations. The $G_{00}$ equation thus reads
\begin{equation}
\frac{3\dot{a}^2}{a^2} - \frac{\beta''}{\beta} - \frac{\chi''}{\chi} - \frac{\beta' \chi'}{\beta \chi}= \kappa \rho a^2. \label{G00.1}
\end{equation}
The $G_{01}$ equation is identically verified.

The $G_{11}$ equation becomes
\begin{equation}
- \frac{2 \ddot{a}}{a} + \frac{\dot{a}^2}{a^2} + \frac{\beta' \chi'}{\beta \chi}= \kappa P_r a^2. \label{G11.1}
\end{equation}

From $G_{22}$, we obtain
\begin{equation}
-\frac{2 \ddot{a}}{a} + \frac{\dot{a}^2}{a^2} + \frac{\chi''}{\chi} = 0. \label{G22.1}
\end{equation}

Finally, $G_{33}$ becomes
\begin{equation}
- \frac{2 \ddot{a}}{a} + \frac{\dot{a}^2}{a^2} + \frac{\beta''}{\beta} = 0. \label{G33.1}
\end{equation}

Now, since we have, in (\ref{G22.1}) and in (\ref{G33.1}), the sum of one part depending only on $t$ and one part depending only on $r$, we can write respectively
\begin{equation}
\frac{\beta''}{\beta} =  \frac{\chi''}{\chi} = c_1^2, \label{1.6}
\end{equation}
\begin{equation}
\frac{2 \ddot{a}}{a} - \frac{\dot{a}^2}{a^2}  = c_1^2, \label{1.7}
\end{equation}
where $c_1$ is a constant, squared for further purpose.
Then, (\ref{1.6}) can be integrated by
\begin{equation}
\beta =  c_3 \textrm{e}^{c_1 r} +  c_4 \textrm{e}^{- c_1 r}, \label{1.8}
\end{equation}
\begin{equation}
\chi =  c_5 \textrm{e}^{c_1 r} +  c_6 \textrm{e}^{- c_1 r}, \label{1.9}
\end{equation}
where $c_3$, $c_4$, $c_5$, and $c_6$ are integration constants. Finally, (\ref{1.7}) can be integrated by
\begin{equation}
a = a_2 \cosh^2 \left[\frac{c_1}{2}(a_1 + t)\right], \label{1.10}
\end{equation}
where $a_1$ and $a_2$ are new integration constants.

Then, the field equations allow us to obtain
\begin{eqnarray}
\rho &=& \frac{c_1^2}{\kappa a_2^2 \cosh^4 \left[\frac{c_1}{2}(a_1 + t)\right]}\left\{3 \tanh^2{\left[\frac{c_1}{2}(a_1 + t)\right]} \right. \nonumber \\
&-& \left. \frac{\left(c_3 \textrm{e}^{c_1 r} -  c_4 \textrm{e}^{- c_1 r}\right)\left(c_5 \textrm{e}^{c_1 r} -  c_6 \textrm{e}^{- c_1 r}\right)}{\left(c_3 \textrm{e}^{c_1 r} +  c_4 \textrm{e}^{- c_1 r}\right)\left(c_5 \textrm{e}^{c_1 r} +  c_6 \textrm{e}^{- c_1 r}\right)} - 2 \right\}, \label{1.11}
\end{eqnarray}
\begin{equation}
P_r = \frac{c_1^2}{\kappa a_2^2 \cosh^4 \left[\frac{c_1}{2}(a_1 + t)\right]}\left[\frac{\left(c_3 \textrm{e}^{c_1 r} -  c_4 \textrm{e}^{- c_1 r}\right)\left(c_5 \textrm{e}^{c_1 r} -  c_6 \textrm{e}^{- c_1 r}\right)}{\left(c_3 \textrm{e}^{c_1 r} +  c_4 \textrm{e}^{- c_1 r}\right)\left(c_5 \textrm{e}^{c_1 r} +  c_6 \textrm{e}^{- c_1 r}\right)} - 1 \right]. \label{1.13}
\end{equation}

\subsubsection{Final form of the class 1 solutions} \label{finalI}

The metric reads
\begin{equation}
\textrm{d}s^2=-A^2 \left[\textrm{d}t^2 - \textrm{d}r^2 - \left(c_3 \textrm{e}^{c_1 r} + c_4 \textrm{e}^{- c_1 r}\right)^2 \textrm{d}z^2 -  \left(c_5 \textrm{e}^{c_1 r} + c_6 \textrm{e}^{- c_1 r}\right)^2\textrm{d}\phi^2\right], \label{metricfin1}
\end{equation}
with
\begin{equation}
A(t) = a_2 \cosh^2 \left[\frac{c_1}{2}(a_1 + t)\right]. \label{Afin1}
\end{equation}

The energy density and the pressure are given by (\ref{1.11}) and (\ref{1.13}), respectively.

\subsubsection{Junction condition} \label{rjunction}

We have already seen that the junction conditions of a cylindrically symmetric source collapsing to a vacuum Einstein-Rosen spacetime have been worked out by Di Prisco et al. \cite{DP09}. The main junction condition setting constraints on the parameters of the interior spacetimes, independently from the exterior parameters, has been established as 
\begin{equation}
P_r \stackrel{\Sigma}{=} 0. \label{J1r}
\end{equation}

Examining $P_r$ as given by (\ref{1.13}), it appears that this condition is indeed realized by solutions belonging to class 1 provided that
\begin{equation}
c_3 c_6 + c_4 c_5 = 0. \label{J2r}
\end{equation}
However, imposing this constraint on the parameters implies that $P_r$ vanishes not only on the boundary, but also everywhere in the interior spacetime. Therefore, the gravitational fluid is a pressureless dust. This case has already been studied in \cite{DP09}.

\subsection{Class 2} \label{class 2}

Class 2 is defined by 
\begin{equation}
\dot{A} = 0 \quad \Rightarrow \quad A = \alpha(r), \label{2.1}
\end{equation}
which we insert into (\ref{shear4}) and obtain
\begin{equation}
\dot{A} = \dot{B} = \dot{C} = 0, \label{2.2}
\end{equation}
which we substitute into the Bianchi identity (\ref{Bianchi3r}) which becomes
\begin{equation}
\dot{\rho} = 0. \label{2.3}
\end{equation}

The solution is therefore stationary and cannot represent a collapse.

\subsection{Class 3} \label{class 3}

Class 3 is defined by
\begin{equation}
A = \lambda \textrm{e}^{b_1 t + b_2 r}, \label{3.1}
\end{equation}
from which we obtain, by inserting it in (\ref{three2r}),
\begin{equation}
2 b_2 \left(\frac{\chi'}{\chi} - \frac{\beta'}{\beta}\right) + \frac{\chi''}{\chi} - \frac{\beta''}{\beta} = 0. \label{3.2}
\end{equation}
A solution of (\ref{3.2}) is
\begin{equation}
\beta = b \textrm{e}^{-2 b_2 r}, \quad \chi = c \textrm{e}^{-2 b_2 r}, \label{3.3}
\end{equation}
where $b$ and $c$ are integration constants. Then, from the field equation (\ref{G11r}), we obtain an expression for the pressure as
\begin{equation}
P_r = - \frac{(b_1^2 + b_2^2)}{\kappa \lambda^2 \textrm{e}^{2(b_1 t + b_2 r)}}. \label{3.4}
\end{equation}
This expression cannot vanish for any finite value of $t$ and $r$, since, if ever $b_1 = b_2 = 0$, this would imply a stationary solution, hence no collapse. The matching condition $P_r =0$ on the boundary $\Sigma$ cannot be satisfied. Therefore, this class 3 is ruled out as a possible solution for our purpose.

\section{Conclusion} \label{concl}

The cylindrical collapse of shearfree anisotropic fluids matched to an exterior ER vacuum has been analyzed here through the use of new exact analytic interior solutions displayed for the first time. This work is a continuation of a previous study by Di Prisco et al. \cite{DP09} where the Robertson-Walker dust solution had been considered and had been shown to lack any radiation transfer to or from the interior.

Here, we have taken anisotropy into account by dividing the problem into three different configurations, each with the pressure along one of the principal directions, i. e., axial, azimuthal and radial.
The three classes of exact solutions found for each pressure configuration have been matched to the ER vacuum exterior. Thus, we have shown that the Darmois matching conditions cannot be satisfied by the fluid with radial pressure, while the axial and azimuthal cases, provided that their lapse function depends only on the time coordinate, do satisfy this constraint.

Then, we have established for both these axial and azimuthal subcases a sufficient condition for an emission of radiation from the interior toward the exterior, This shows that, while an isolated spherically symmetric collapsing object cannot emit gravitational waves, cylindrical ones with axial or azimuthal pressure can, under the conditions displayed in this paper. Indeed,  in the realm of GR, spherical gravitational waves are forbidden by Birkhoff's theorem, while this theorem does not apply to the case of cylindrical symmetry.

Of course, the shearfree assumption retained here is only valid for a slowly collapsing object. It might break down at the end of the collapse when the pressure fails to counteract gravitation sufficiently.

However, should this simplifying assumption be given up, analytic solutions such as the ones used here could arise no more. Anyhow, the results displayed here can increase our understanding of gravitational radiation, by showing that, even in the simplified picture of an infinite cylinder of anisotropic shearfree matter, gravitational waves can be emitted during collapsing motion.

\end{document}